# Legal Provocations for HCI in the Design and Development of Trustworthy Autonomous Systems

Lachlan D. Urquhart*

School of Law, University of Edinburgh. lachlan.urquhart@ed.ac.uk

Glenn McGarry & Andy Crabtree

School of Computer Science, University of Nottingham. glenn.mcgarry@nottingham.ac.uk, andy.crabtree@nottingham.ac.uk

We consider a series of legal provocations emerging from the proposed European Union AI Act 2021 (AIA) and how they open up new possibilities for HCI in the design and development of trustworthy autonomous systems. The AIA continues the 'by design' trend seen in recent EU regulation of emerging technologies. The AIA targets AI developments that pose risks to society and citizens' fundamental rights, introducing mandatory design and development requirements for high-risk AI systems (HRAIS). These requirements regulate different stages of the AI development cycle including ensuring data quality and governance strategies, mandating testing of systems, ensuring appropriate risk management, designing for human oversight, and creating technical documentation. These requirements open up new opportunities for HCI that reach beyond established concerns with the ethics and explainability of AI and situate AI development in human-centered processes and methods of design to enable compliance with regulation and foster societal trust in AI.

> *"The proposal is without prejudice and complements the General Data Protection Regulation and the Law Enforcement Directive with a set of harmonised rules applicable to the <u>design, development and use of certain high-risk AI systems</u>"* (European Commission, Proposal for an AI Act).

## 1 INTRODUCTION

We consider proposed European regulations targeted at autonomous systems, particularly the recent proposal for an AI Act (AIA) [18] and its implications for HCI. The AIA continues the bold move made by the General Data Protection Regulation (GDPR) [16], which introduced legal requirements in 2016 to "implement appropriate technical and organisational measures" enabling "data protection by design and default" (GDPR, Article 25). This introduced a unique set of legally mandated *design requirements* to bridge between legal principles and the technologically-driven use of personal data in the EU and beyond. The AIA continues this trend towards 'by design' regulation, introducing "common mandatory requirements" for the "design and development" of AI systems (AIA, Explanatory Memorandum). The overarching aim in both cases is to create *trust*. For GDPR, the issue of trust lay in fostering consumer confidence in the digital economy (GDPR, Recital 7); for the AIA, the issue is to "ensure AI is developed in ways that respect people's rights and earn their trust" (AIA, Explanatory Memorandum). Through design and development, the AIA thus seeks to "promote trustworthy AI" and create an "ecosystem of trust" (ibid.).

---

* Place the footnote text for the author (if applicable) here.

Our goal in this paper is to sensitise the HCI community to the mandatory design and development requirements of the AIA. Yet to be adopted and implemented in law, the AIA nevertheless aims to shape "global norms and standards" in AI regulation (AIA, Explanatory Memorandum). In doing so it creates a series of *legal provocations* for AI that open up new opportunities for HCI beyond the established design-oriented concerns with ethics, fairness, bias, model interpretability and explanation. Instead, the provocations provided by the AIA surface new legal expectations for the design and development of AI and extend our understanding of what having a 'human in the loop' for AI oversight means. Below we unpack these provocations to help develop a shared understanding of the intersection between the AIA and HCI and to consider how this alignment might shape the emergence of trustworthy autonomous systems in everyday life.

We begin by considering the scope, applicability, and operational context of the AIA. We then unpack the legal conception of trust the regulation trades on before turning to a selection of mandatory design and development requirements imposed on AI systems. These requirements seek to build societal trust in AI by mandating data governance practices, testing regimes, risk management, human oversight, and stringent technical documentation. We move on to consider how these legal provocations align with and open up new opportunities for HCI. This includes finding a greater role for HCI in the technical heartlands of AI development, exploring novel means of rendering AI systems accountable, developing human-machine interface tools for better oversight and action in monitoring HRAIS, and embedding AI system development in human-centred design methods and processes, as well as legal processes, to enable compliance. The legal provocations encapsulated in the AIA show that there is much more to the "human side" of AI [1] from a regulatory perspective than just focusing on the interpretability of machine learning systems and explainable AI and the commensurate need to make AI systems intelligible to ordinary users. This paper thus sets out a novel agenda that aligns HCI with the regulation of AI and situates the design and development of AI systems in human-centred approaches by *fiat*.

## 2 INTRODUCING THE PROPOSED EU AI ACT 2021

The AIA defines AI systems as those that use machine learning approaches, including supervised, unsupervised, reinforcement, and deep learning approaches, or logic and knowledge-based approaches, or statistical including Bayesian approaches (Annex I, AIA). The AIA classifies AI systems in terms of the differentiated *risk* they present to citizens and society. At the base of the pyramid is *low* risk AI, systems that pose minimal risks and on which fewer compliance obligations (e.g., transparency) are placed. In the middle are *high* risk AI systems, which face wide ranging design and development requirements. And at the top of the pyramid is *prohibited* AI, which covers systems that pose risks that are unacceptable to health, safety, and citizens fundamental rights (AIA, Article 5).[1] We primarily focus on the regulation of high-risk AI systems (HRAIS) in this paper, as they are subject to comprehensive 'by design' regulation measures. An AI system is deemed high-risk in the AIA through one of two mechanisms. The first is if *both* of the following conditions are fulfilled: 1) an AI system is either a product itself or is intended to be used as a safety component of a product covered by the EU harmonisation legislation listed in the annexes to the AIA (Article 6) *and* 2) the product or AI safety component is required to undergo a third-party conformity assessment. The Annexes to the AIA [19] provide a long list of related legislation covering a broad range of riskier product types including toys, machinery, lifts, fuel burning appliances, protective equipment, radio equipment, medical devices, civil aviation, motorised vehicles, watercraft and marine equipment and rail in which AI could be used and are thus brought under the remit of the AIA (Annex II). Secondly, an AI is high risk if it is used in one of the contexts listed in Annex III. Application domains include the use of AI in critical national infrastructure; educational and vocational training; employment, including worker management and access to self-employment; access to and enjoyment of essential public or private services and benefits; law enforcement, migration, asylum and border control; administration of justice and democratic process; and biometric identification and categorisation uses (Annex III). There is a process for updating this list over time by the European Commission, providing some degree of future proofing. The regulation stipulates that a *quality management system* (QMS), already commonplace in business,[2] must be put in place by the provider of a HRAIS to manage the compliance requirements and risks that a system poses across these product areas and application domains. The QMS may be subject to "internal control" (self-assessment) or third-party conformity assessment by a "notified body" designated by a member state to do conformity assessments [21].[3] The AIA also requires a declaration of EU conformity be drawn up by the provider (Article 48 and Annex V), and a CE mark of conformity affixed "visibly, legibly and indelibly" (Article 49) to packaging and documentation before a high-risk system is put on the EU market.

---

[1] For example, the AIA places prohibitions on certain uses of remote biometric identification systems (facial recognition). Thus, and with certain caveats, the "use of 'real-time' remote biometric identification systems in publicly accessible spaces for the purpose of law enforcement is prohibited, unless it is strictly necessary" (i.e., unless occasioned by a threat to public safety or the apprehension of persons involved in significant criminal offences). Prohibited too is the use of AI systems that deploy subliminal techniques, that exploit the vulnerabilities of persons in order to materially distort behaviour in a manner that causes or is likely to cause physical or psychological harm, and the use of social scoring systems by public authorities to evaluate and/or classify the trustworthiness of people based on their social behaviour or known or predicted personal or personality characteristics (AIA, Article 5). For clarification on the caveats to facial recognition and the wider use of AI by law enforcement, see Urquhart and Miranda [64].

The QMS is key to the management of risk and must furnish an account of the provider's compliance strategy, technical documentation, and the procedures put in place to ensure the efficacy of the QMS over a system's lifetime [19]. When third-party conformity assessment is required, this may include providing the designated body with 'full access to training and testing datasets' through an application programming interface (API) or other means of remote access. Alternatively, the provider of the system may be obliged to provide 'access to the premises where the design, development, testing of the AI systems is taking place'. Upon "reasoned request" (AIA, Annex VII), the notified body must also be granted access to the source code of the AI system and may require that the provider supply further evidence or carry out further tests to enable a proper assessment of conformity. If and when an AI system is deemed to be in conformity, an EU technical documentation assessment certificate is issued by the designated body, who is also tasked with conducting periodic audits to ensure the provider maintains and continues to apply the QMS. Any change or modification to the system that could affect its compliance or its intended purpose must be approved by the designated body which issued the assessment certificate [19] (and must otherwise be subject to internal control). It is notable that systems that continue to learn after being placed on the market or put into service "shall not constitute a substantial modification" (AIA, Article 43).

It is notable too that the 'provider' of an AI system is not necessarily the developer or manufacturer of an AI system. The AIA thus *distributes responsibility* for compliance across the supply chain.

> "Any distributor, importer, user or other third-party shall be considered a *provider* for the purposes of this regulation and shall be subject to the obligations of the provider ... if any of the following circumstances apply:

---

[2] See, for example ISO 9001 [32].
[3] The need for third-party conformity assessment turns on various considerations including whether or not harmonised legislation already provides mechanisms for assessing conformity, or "harmonised standards" published in the *Official Journal of the European Union* have been applied (AIA, Article 40) or "common specifications" adopted through future EU legislation have been applied (AIA, Article 41). If the answer is no, then third-party conformity applies unless a provider is eligible for "internal control" (AIA, Article 43) or can follow the compliance process within the harmonisation legislation (AIA, Article 43). The rules also differ for application domains covered by Annexes II and III, according to Article 43 (AIA) e.g., it differs for HRAIS conducting biometric identification and categorisation.

> a) they *place on the market* or put into service a high-risk AI system under *their name or trademark*; b) they *modify the intended purpose* of a high-risk system already placed on the market or put into service; or c) they make a *substantial modification* to a high-risk system" (AIA, Article 28, our emphasis).

The distribution of responsibility imposes legal obligations on importers to ensure a conformity assessment has been carried out by the provider (AIA, Article 26). If importers cannot be identified, providers based outside the EU must appoint an authorised representative in the EU prior to making their systems available on the EU market (AIA, Article 25). Distributors must verify that high-risk AI systems comply with the obligations set out in the regulation (AIA, Article 27). Users must inform the provider or distributor if they identify any serious incidents or malfunctioning and interrupt the use of the system in such circumstances, and they must suspend the use of the system if they have reason to consider it presents a risk to human health, safety and the protection of fundamental rights (AIA, Article 29).

These measures are designed to create an "ecosystem of trust in AI in Europe" by putting in place a legal framework "centred on a well-defined risk-based regulatory approach" that "imposes regulatory burdens only when an AI system is likely to pose high risks to fundamental rights and safety" (Explanatory Memorandum, AIA). This is seen as "a better option than blanket regulation of all AI systems" (ibid.). Whilst the regulation covers providers and users located in the EU, it will, like GDPR, have global reach, and includes mechanisms to expand its scope beyond the EU. For example, providers who put AI systems on the market or into service in the EU, whether they are established in the EU or third countries, will be subject to these rules. Also, where there are users or providers of AI who are based in a third country but the outputs of the AI system are utilised in the EU, they will again be covered by the rules (AIA, Article 2). Furthermore, like, GDPR, failure to comply with the regulation will be subject to significant administrative fines, this time up to 30 million euros or 6% of total worldwide annual turnover, whichever is higher. These measures, including the introduction of a risk-based regulatory approach, third party conformity assessment of quality management systems, the distribution of responsibility across the supply chain, and global reach are seen as key to mechanisms for promoting "trustworthy AI."[4]

## 3 THE LEGAL CONCEPT OF TRUSTWORTHINESS

Before we move on to consider the specific design and development requirements mandated by the AIA, we wish to briefly consider the legal notion of trustworthiness that frames the proposed regulation. The legal notion is related to and informed by but not the same as or reducible to ethical conceptions of trustworthiness, which have informed technical agendas in AI. An ethical framing of trustworthiness is encapsulated in the High-Level Expert Group (HLEG) on Artificial Intelligence report *Ethics Guidelines for Trustworthy AI* [31]. The HLEG brought together a self-nominated group of experts from industry and academia to consider the principal ethical challenges confronting the widespread uptake of AI in society [65] and subsequently they defined trustworthy AI as having 3 essential components. That AI is *lawful* and complies with fundamental rights including the freedom of the individual; respect for human dignity; citizens' rights; equality, non-discrimination and solidarity; and respect for democracy, justice and the rule of law. That AI is *ethical* and respects 4 fundamental principles including respect for human autonomy; the prevention of harm to human beings; fairness and ensuring equality of opportunity and the right to redress; and explicability or the need to be transparent about AI decision-making. That AI is *robust* from a social and environmental as well as technical perspective, not only being well engineered but also being fit for purpose within the context in which a system will be deployed and operate [31]. These components create a complex framing of trustworthiness where multiple and even competing values are at play and create distinct challenges for design [24].

---

[4] One immediate consequence, and indeed some may see it as positive benefit, is that many emerging applications of AI in society are likely to be seen and treated as high-risk. However, there is concern about whether or not the net should be more widely cast. For example, commercial uses of emotion sensing systems are currently low risk [38], yet regulatory bodies such as the EU Data Protection Board and EU Data Protection Supervisor have called for these systems to be banned [22].

From the point of view of robustness there is strong alignment with *trustworthy computing* [26] and that systems are therefore secure, respect privacy, are reliable, have business integrity, are thus used in responsible ways [43], and align with the earlier notion of *dependable computing*. As Randell [47] puts it,

> " … the trustworthiness of a computer system [is] such that reliance can justifiably be placed on the service it delivers. Dependability thus includes such properties as reliability, integrity, privacy, safety, security, etc., and provides a convenient means of subsuming these various concerns within a single conceptual framework."

These concerns, which are far from new or novel and in no way confined to AI, are writ large in the HLEG report and are closely linked to the ethical principle of preventing harm. Robustness, and especially technical robustness is closely aligned then with human safety and leads to consideration of further AI-specific issues including the resilience of AI systems to attack and the need to implement fall back plans to minimise unintended consequences and errors. Trustworthiness thus comes to have a particular framing from the point of view of robustness that emphasises the engineering of "safe, secure and reliable" systems that "prevent any unintended adverse impacts" [31].

It is notable, that Mundie, author of Microsoft's *White Paper on Trustworthy Computing* reminds us, that "making something trustworthy requires a social infrastructure as well as solid engineering" [43]. It is perhaps in this context that the ethical framing of trustworthiness is best understood. This certainly applies to the law, to which ethics in AI is closely bound (but not isomorphic). As the HLEG report puts it, its ethical principles are "rooted in fundamental rights", which are "to a large extent already reflected in existing legal requirements" but go "beyond formal compliance with existing laws" [31]. Indeed, they do, bleeding quite recognisably into technical agendas in AI. Thus, in building on the principles outlined in the HLEG report, it is not only recommended that trustworthy AI incorporate *human oversight* to ensure that AI systems do not undermine human autonomy, but also that robust *data governance* be implemented to control access to personal data, ensure the quality and integrity of data, and protect data subjects' privacy. It is recommended too that the process and outcome of automated decision-making be *transparent*, which includes explaining AI processes and outcomes, tracing and auditing decisions, and clearly communicating AI capabilities and limitations. AI systems should respect diversity and be non-discriminatory and thus *fair*, which involves tackling algorithmic bias, and they should be *accountable* both in their development and use, particularly to "public enforcers". Trustworthiness thus comes to have a particular framing from the point of view of ethics that emphasises the engineering of a principled social infrastructure that intersects with and informs AI design and development. This intersection is often characterised by the 'FAccT' (Fairness, Accountability and Transparency) agenda, which sees the technical community trying to develop ways and means of tackling principled ethical concerns [e.g., 35, 51, 52, 53].

However, fairness, accountability, and transparency are not only ethical and technical concepts[5]; they are key to *good governance* which requires governments and industry treat people equally without discrimination, ensure ongoing answerability for decision-making, and provide openness around decisions [9]. Thus, whilst there is a need to discuss ethical and technical dimensions of building trustworthy AI, *policy and law* are key mechanisms for enabling governance [7] and the AIA makes it clear that trustworthiness in the eyes of EU governance is fundamentally about **risk management**. The AIA configures trustworthiness as the management through legislation of risks to human health, safety, and fundamental rights occasioned by algorithmic discrimination in the use of digital services and the use of AI in safety components of all manner of products, devices, machinery and machines including large-scale IT systems (AIA, Explanatory Memorandum). The idea that trustworthiness is a matter of risk management will be familiar to many in HCI, where risk has previously been considered as a *personal* disposition, attitude, or belief grounded in rational choice and the calculus of risk [25, 8]. However, the legal concept has more in common with *societal* arrangements of trust, which locate trustworthiness in such things as the social contract, social capital, and social institutions [59, 46] amongst other mechanisms of social order [40]. Thus, while informed by ethical conceptions and related technical developments, the legal notion of trustworthiness is distinct from these framings. The AIA creates a formal legal basis mandating AI providers take certain actions to manage the risks posed AI systems backed by the authority of law. What constitutes trustworthiness is no longer a matter of personal or professional opinion but ineluctably defined in law. The AIA cannot be dismissed or ignored, then, and will impact 'ethics washing' [69] where AI providers have sought to avoid regulation by ascribing to high level principles and adopting ethical codes of practice without meaningfully tackling actual or potential social harms.

---

[5] These concepts are legal principles and frameworks too e.g., the GDPR Art 5(2) Accountability Principle.

## 4 MANDATORY DESIGN AND DEVELOPMENT REQUIREMENTS

While the AIA implements a legally-binding version of what constitutes and provides for trustworthiness it is not, as we have said, insensitive to ethical or technical framings. Indeed, the AIA is remarkable in the degree of granularity it adopts in setting out mandatory design and development requirements for high-risk AI systems (Chapter 2), which have clearly been informed by ethical concerns. For example, algorithmic bias and its consequences have been one of the primary concerns amongst technical and ethical communities around emergence of AI systems. It takes centre stage in the proposed regulation too:

> " … the proposal complements existing Union law on non-discrimination with specific requirements that aim to minimise the risk of algorithmic discrimination, in particular in relation to the design and the quality of data sets used for the development of AI systems" (AIA, Explanatory Memorandum).

The AIA thus requires HRAIS that make use of "techniques involving the training of models with data" to comply with "**quality criteria**" laid down in the proposed regulation. Thus designers are legally obliged to develop training, validation, and testing data sets to legally set standards. In doing so designers must ensure that suitable data sets are available and in sufficient quantity to meet the intended purpose of the system. They need to ensure data sets are representative and have appropriate statistical properties that take specific geographical, behavioural, or functional characteristics of the setting where the HRAIS is to be deployed (e.g., training data reflects the population that will be subject to the system). They also need to make sure data sets are free of errors and that any shortcomings, including gaps in the data, are identified along with how to address them. Lastly, data sets should be examined for possible biases, and where training data sets are concerned, all of this should be done prior to model development (AIA, Article 10).[6] Designers must subsequently measure levels of accuracy and publish accuracy metrics in the instructions of use, and develop feedback loop based mitigation measures for systems that continue to learn after being put into service, to circumvent emergent bias (AIA, Article 15).[7]

The focus on improving quality of training datasets and preventing discrimination by mandating common data governance practices is "complemented with obligations for **testing, risk management, documentation and human oversight** throughout [an] AI systems' lifecycle" (AIA, Explanatory Memorandum). In addition to model development,

---

[6] It has been noted that these mandatory data governance practices are not sufficient and need to be further bolstered through regulation to take into account "techniques such as unsupervised or reinforcement learning [which] do not rely on validation and testing data sets" [58].

[7] It has been noted too that accuracy is not the only measure of model performance, e.g., "One of the performance metrics used in reinforcement learning is its reliability. The requirements in Article 15 should accommodate such performance metrics" (ibid.).

HRAIS must be tested prior to being put on the market to ensure they perform consistently against "preliminarily defined metrics and probabilistic thresholds that are appropriate to the intended purpose" (AIA, Article 9). More radically, HRAIS must also be tested "for the purposes of identifying the most appropriate risk management measures" and this includes identifying foreseeable risks when a system is used according to its intended purpose, and risks that may arise under conditions of *reasonably foreseeable misuse* (AIA, Article 9). The latter includes testing that systems are resilient to attempts by unauthorised third parties to alter their use or performance, including cyberattacks that try to manipulate training data ('data poisoning') and inputs designed to cause the model to make a mistake ('adversarial examples') (AIA, Article 15). The reasonably foreseeable misuse clause puts the systematic use of adversarial testing centre stage in the development of HRAIS. As the UK's National Cyber Security Centre reminds us, this is not just a matter of understanding and analysing the risks to data models but risks to the system as a whole occasioned by its entire array of external dependencies [45].[8] The reasonably foreseeable misuse clause represents a significant provocation to the design and development of trustworthy AI, to think beyond intended applications, to anticipate what might trigger these, when, and to design appropriate mitigations and management measures.

Significant too is the legal obligation that the designers of HRAIS put a **risk management system** (RMS) in place prior to a high-risk system being put on the market and that the RMS be updated and maintained throughout the system's lifetime (AIA, Article 9). The RMS requires designers to "take account of the state of the art" and implement technical measures to eliminate risks to human health, safety, or fundamental rights. Alternatively, if risks cannot be wholly eliminated, then risks should be reduced through the implementation of "mitigation and control measures" and any "residual risks" should be clearly communicated to users. Furthermore, the RMS should evaluate risks emerging over the course of a high-risk system's lifetime by putting a **post-market monitoring system** (PMS) in place (AIA, Article 9). The PMS must "systematically collect, document and analyse data" that is provided by users or collected through other sources "on the performance of a system throughout its lifetime" (AIA, Article 61). Complementing this, the AI Act requires that HRAIS "be designed and developed with capabilities enabling the automatic recording of events ('logs') while the system is operating" (AIA, Article 12). Logging "should facilitate post-market monitoring" (ibid.) and thus enable the provider to monitor the continuous compliance of the system and situations that may result in the system posing risks to health, safety or fundamental rights or that occasion the need for substantial modification.

The RMS and PMS are part and parcel of the QMS and must be described in accompanying **technical documentation**, which should again be drawn up before a high-risk system is put on the market or into service (AIA, Article 11 and Annex IV). Technical documentation should minimally provide a *general description* of the system (including intended purpose, hardware requirements, interaction with external software and hardware, versions and update requirements, a schematic of the product in which the system is a component if applicable; installation and user instructions). This should be accompanied by a *detailed description* of the system and the process for its development including *design specifications* (key design choices, assumptions about the data and data subjects, logic of the AI and its algorithms, main classification choices, optimisation choices and trade-offs); *system architecture* (including integration of software components); *data requirements* (datasheets describing data selection and provenance, characteristics of the data, training methodologies, labelling procedures, and cleaning methodologies); *training, testing and validation procedures* (including tests logs, metrics used to measure accuracy, robustness, cybersecurity and compliance, and assessment of discriminatory impacts); and *assessment of human oversight measures* (including technical measures to facilitate interpretation of outputs). Detailed information about the *monitoring, functioning and control* of the system should also be provided (including its capabilities and performance limitations, accuracy measures, foreseeable unintended outcomes and sources of risk to health, safety, fundamental rights and discrimination, and measures needed to enable human oversight), along with a detailed description of the *RMS* and *PMS*, any *changes* made to the system, and the harmonised EU *standards* that apply to the system (AIA, Article 11 & Annex IV).

---

[8] It is worth noting, as found in a recent study, that popular off-the-shelf machine learning frameworks, including Caffe, TensorFlow and Torch "contain heavy dependencies on numerous open source packages." Indeed, "Caffe is based on more than 130 depending libraries .. and Tensorflow and Torch depend on 97 Python modules and 48 Lua modules respectively [72].

Monitoring, functioning, and control requirements are not only matters of documentation. Designers will both have to incorporate elements of the RMS into the development process (e.g., data governance and testing) and develop and implement elements of the RMS and PMS as technical components within HRAIS (e.g., state of the art risk mitigation measures and automatic logging). The proposed regulation will also oblige designers to build-in measures that enable people to exercise control over HRAIS:

> "High-risk AI systems shall be designed and developed to ensure that their operation is sufficiently transparent to enable users to interpret the system's output and use it appropriately with a view to achieving compliance with the relevant obligations of the user and of the provider … " (AI Act, Article 13)

Relevant obligations include and require that, if feasible, **human oversight measures** be identified and built into HRAIS by the provider before a system is placed on the market or that they be identified by the provider and implemented by the user if it is appropriate to do so (AIA, Article 14). This includes the development of contextually or situationally appropriate human-machine interface tools that enable effective oversight during the period in which a system is in use, and which are designed with the express aim of preventing or minimising risks to health, safety or fundamental rights whether a system is used for its intended purpose or under conditions of foreseeable misuse.

The proposed regulation lays down a raft of measures to enable "natural persons" or individuals to whom human oversight may be assigned to do the following as appropriate to circumstance:

> "(a) fully understand the capacities and limitations of the high-risk AI system and be able to duly monitor its operation, so that signs of *anomalies, dysfunctions and unexpected performance* can be detected and addressed as soon as possible;
>
> (b) remain aware of the possible *tendency of automatically relying or over-relying on the output* produced by a high-risk AI system ('automation bias'), in particular for high-risk AI systems used to provide information or recommendations for decisions to be taken by natural persons;
>
> (c) be able to *correctly interpret* the high-risk AI system's output, taking into account in particular the characteristics of the system and the interpretation tools and methods available;
>
> (d) be able to decide, in any particular situation, *not to use* the high-risk AI system or otherwise *disregard, override or reverse the output* of the high-risk AI system;
>
> (e) be able to intervene on the operation of the high-risk AI system or interrupt the system *through a "stop" button* or a similar procedure." (ibid. – our emphasis)

Article 14 also states that human oversight of biometric systems used for remote identification should include verification and confirmation by *at least two natural persons*.[9] Generally, it is expected that HRAIS should be responsive to human operators, be designed and developed so that people can oversee their function and, as appropriate to context and risk, operational constraints should be built-in to ensure HRAIS "cannot be overridden by the systems itself" (Recital 48). This expectation is seen as key to creating trustworthy AI, which "can only be achieved by ensuring an appropriate involvement by human beings in relation to high-risk AI applications" [17].

---

[9] Both in real time and post data collection.

To summarise, common mandatory design and development requirements of the proposed regulation thus frame trustworthiness of HRAIS around the design and development of key requirements. These include:

- Data sets that are in compliance with *quality criteria* and *data governance practices* that regulate the acquisition, training, validation and testing of data sets and commensurate development of models to minimise the risk of algorithmic discrimination.
- The *testing of systems* for accuracy, consistent performance with respect to intended use, and identification of risks, including risks occasioned by reasonably foreseeable misuse of the data, the model or the system as a whole.
- The design and development of *state of the art risk management measures* that eliminate risks to human health, safety or fundamental rights wherever possible, or put mitigation and control measures in place if risks cannot be entirely eliminated.
- The design and development of *automatic logging* that records data on the performance of a system throughout its lifetime and enables post-market monitoring of a system's compliance, the occurrence of situations that pose risks to health, safety or fundamental rights or that otherwise occasion the need for substantial modification.
- The design and development of *human-machine interface tools* enabling users to exercise oversight and control in systems use.

Furthermore, design and development is to be described in *technical documentation* contained in the *quality management system*, which provides a general description of the system, a detailed description of its design specification, system architecture, data requirements, training, testing and validation procedures, and assessment of human oversight measures. Technical documentation must also describe system monitoring, functioning and control, the risk management system and post-market monitoring system, system modifications or changes, and the EU standards that apply, all of which along with training and testing data sets and source code are *subject to third party conformity assessment and certification* as is necessary to the regulation of high-risk AI systems in society.

## 5 TRUSTWORTHY AI & HCI

In its framing of trustworthy AI, the AIA presents a number of significant provocations for design and development. These focus on data governance, testing, risk management, human oversight and technical documentation. Each of these areas presents opportunities for HCI, some perhaps more obviously than others. There has, for example, been longstanding interest in HCI in information management [44] and, more recently, concern with Human Data Interaction has emerged in the effort to put people at the centre of the flow of data in society [41]. The role of HDI in human-centred data governance is, as yet, under-explored particularly in the context of federated machine learning systems [29, 42]. While research has investigated how adversarial testing can be used to disrupt human-computer action [27], the role of HCI in adversarial testing is less pronounced though the possibility clearly exists. As Shneiderman [57] points out, a key part of the testing process, particularly the verification process, is to "develop test scenarios to detect adversarial attacks, which would prevent malicious use." Given this, it would seem evident too that HCI may have a significant role to play in risk management, both prospectively and retrospectively. Prospectively, there is long tradition of risk assessment in the human factors literature spanning a broad range of domains, which includes methods for identifying and assessing risks as part of the process of usability evaluation prior to systems deployment [e.g., 6, 33, 50, 70, etc.]. The prospective identification and assessment of risk in AI systems would appear to be a strong area for continued HCI research and methodological development. Retrospectively, there is also a long tradition of studying systems deployed in the wild through fieldwork and this approach may prove useful in post-market monitoring of HRAIS. For example, a recent study of the deployment of AI to detect diabetic retinopathy [2] revealed the impact of workflow, infrastructure, and culture on the performance of the system and its ability to actually detect diabetic retinopathy. HCI may, then, play a formative role in managing the risks that foreseeably accompany the development of AI systems and those that otherwise emerge after their introduction into actual settings of use.[10]

There may also be a role for HCI to play in the technical documentation of AI systems. This may not seem particularly obvious or attractive. However, the issue here concerns the role played by technical documentation in regulation, which is to demonstrate the compliance of a system with the AIA. The demonstration particularly concerns Articles 8 to 15 which include data governance, risk management, human oversight, and cybersecurity and is a legal obligation which, similar to the *accountability* principle enshrined in GDPR (Article 5(2)), is key to showing that a provider is in compliance with the law. Declaration is not sufficient; a provider must be able to demonstrate compliance on demand even if they self-certify. Previous work at the interface between law, technology and HCI has explored how accountability might take different forms, being demonstrated, for example, through a human-centred architecture that communicates features of a system and how they map onto different GDPR requirements [63]. Accountability in GDPR covers various cohorts including the data controller and data protection officer, the data subject, and supervisory authorities. The AIA only requires accountability to supervisory authorities (notified bodies and regulators) who, like the provider, have a technical interest in determining the compliance of HRAIS. There may, then, be a significant role for HCI that extends beyond literal documentation to shape the architectures of human-centred AI systems in collaboration with developers and technology law scholars to enable novel demonstrations of compliance. There will be no one-size fits all human-centred AI architecture but many and this would be a substantial strand of research in its own right, cutting to the heart of what it is to *make* AI human-centred across different classes of autonomous system embedded within diverse application domains subject to various regulations.

We save **human oversight** until last as it is the most obvious point of connection between AI, its regulation, and HCI. The requirement that HRAIS "be designed and developed to ensure that their operation is sufficiently transparent to enable users to interpret the system's output and use it appropriately" (AIA, Article 13) aligns strongly, on the face of it, with ethical concerns with explicability and the perceived need for transparency in AI decision-making to engender trust. This bleeds into two closely related technical agendas: interpretability in machine learning (IML) [36] and explainable AI (XAI) [62]. If there is a difference between the two it might be that IML has a broader focus than XAI. IML is of longstanding concern, reaching back at least 30 years and spans various "objectives" including understanding the causal nature of automated decision-making, the transferability and generalisation of models and outputs, their ability to inform human decision-making and assessing whether or not automated decisions are fair as well as trustworthy [36]. XAI, on the other hand, emerged more recently and does not seem to cover quite the same ground focusing, for example, on "the machine's current inability to explain their decisions and actions to human users" [62]. Transparency, like interpretability and explainability, is multi-faceted and implicates models, their individual components, and algorithms [36]. As Abdul et al. [1] note in their extensive literature review of papers on explainability,

---

[10] Revisions to the EU Machinery Directive 2006 [20] also extend HCI's role in managing risks to include those created by collaborative robots and driverless mobile machinery, which chimes strongly with HCI's ongoing interest in human-robot interaction [49].

> "This has produced a myriad of algorithmic and mathematical methods to explain the inner workings of machine learning models ... However, despite their mathematical rigor, these works suffer from a lack of usability, practical interpretability and efficacy on real users."

The challenge of ensuring that the operation of AI systems is sufficiently transparent to enable users to interpret their outputs and use them appropriately is underscored by other researchers too, who found in a recent study of machine learning deployments that explainability methods are, for the most part, only used by engineers to debug machine learning models, improve their performance, and to sanity check their work [3]. There is then, as Abdul et al. [1] point out, a pressing need for the HCI community to take the lead and ensure AI systems do not ignore the "human side" of AI and thus ensure that systems are explainable to their users as well as those who develop and maintain them, as will soon be required by law.

**5.1 The Human Side of AI**

If HCI is to take a lead and help AI meet its legal design and development obligations, it is important that attention is paid to the fine detail of the proposed regulation. The explainable AI agenda stemmed in part, for example, from the technical community's (mis)reading of GDPR, which was seen to mandate a "right to an explanation" of automated decision-making using personal data [28]. This led to debates about how to make automated decision-making explainable to citizens and spurred the growth of new techniques, including non-machine based techniques such as counterfactuals [68], in a bid to help people understand how AI systems arrive at decisions. Despite an active technical agenda in XAI, legal debate has continued to discuss *if* a right to an explanation exists or not in GDPR, its breadth, on what legal basis it might exist, and if it is even the best remedy to protect user interests and build trust in autonomous systems [67, 54, 14, 55, 12, 11]. As these discussions show, the law is interpretative and rarely provides clean rules that tell us what the 'correct' response from designers should be. The shift to 'by design' regulation is an attempt to bridge this gap, though we still need to be sensitive to the fine print. It thus becomes apparent that the AIA situates transparency and interpretation *in* human oversight:

> "High-risk AI systems shall be designed and developed in such a way, including with appropriate human-machine interface tools, that they can be effectively overseen by natural persons during the period in which the AI system is in use." (AIA, Article 14)

It is important to recognise the situated nature of transparency and interpretation. It contextualises the meaning of the Article 13 transparency and interpretation requirement and what the raft of measures it specifies should enable. Seen from within the context of human oversight, explanation is not framed as a guarantor of trust then, but is part of a package of human-machine interface measures that are to be designed and developed to support *action*. It is not action (or interaction) with the machine per se, but action in the context of machine use. Thus, what the AIA requires by way of transparency and interpretation is that human interface tools support human oversight and thus enable human operators to take action to prevent or minimise risks to human health, safety or fundamental rights occasioned either by intentional use or reasonably foreseeable misuse of a high-risk AI system. **Action not explanation is imperative**; transparency and interpretation are subservient to this goal.

The transparency and interpretability requirement of the AIA should not be taken at face value then and understood to segue neatly with ethical principles, IML, or XAI methods, or the need to extend them beyond developers to actual users. On the contrary, the mandate is not an invitation to dive down the rabbit hole of interpretability, but to design and develop human-machine interface tools that enable users to take action to prevent or minimise actual or potential harms. Explanation has a role to play in this achievement, but there is much more to it than explanation can provide. While there is, as Abdul et al. [1] point out, both scope and need to help AI develop methods of interpretation and explanation that are usable by non-developers in the course of their practical activities, the legal requirement for human oversight opens up the design space. The emphasis on explanation might be complemented by the design of *intelligent user interfaces* [66] and human-machine interface tools that support *collaborative decision-making* [73]. These approaches have been leveraged to bridge between AI and HCI in the effort to support interpretation and explanation and build the human into the decision-making loop. Both approaches also face fundamental challenges which must be addressed if they are to get to grips with the human oversight requirement. New design frameworks are required to strike a balance between excessive automation and excessive human control [56], for example, and there is pressing need to address the fundamental challenge of aligning collaborative solutions with situated human reasoning [5]. However, both the legal and practical need to build interfaces usable by human operators into AI systems speaks to further HCI expertise.

Of particular relevance is HCI expertise in understanding the context of use in which systems are deployed and shaping computing systems (and AI systems *are* just computing systems, despite the hype) around users to support situations of use.

> "High-risk AI systems should be designed and developed in such a way … [as to be] responsive to the human operator … the natural persons to whom human oversight has been assigned [should] have the necessary competence, training and authority to carry out that role." (AIA, Recital 48)

This recital, which provides interpretative guidance, may be read literally as placing certain obligations on human operators or it may be understood to reflect the fundamental requirement that human-machine interface tools should be designed and developed to be responsive to user competence and training. Not competence in using an AI system per se, though this is clearly part of the mix (AIA, Article 29), but competence in using the system for some *purpose*. This recognises that AI systems have an indexical relationship to and will thus be embedded in organised settings of use. They therefore need to be designed and developed to support human oversight with respect to those organised settings of use and what persons therein will *do* with them. As Harper puts it in describing the role of HCI in the age of AI, *"doings with AI* are more important than the *mechanics of AI"* [30].

Designing human-machine interface tools to support situated "doings with AI" (including doing oversight) gives rise to the need to develop what Harper calls "new grammars of action" to mediate human-machine interaction (ibid.). It is an idea that has previously been used to characterise interdisciplinary design approaches that leverage Wittgensteinian thinking and orients HCI researchers to ordinary language concepts as resources for the study of organised settings and the design of technologies to support the everyday activities which articulate those concepts [e.g., 61, 10, 15]. In more prosaic terms, understanding and designing for grammars of action involves ethnographic study and cooperative prototyping, both of which are commonplace in HCI along with an arsenal of other methods for understanding users, the contexts in which they operate, their organisational needs and for tailoring technological support accordingly through the active participation of users in the design process. The need to take account of users, the contexts in which they operate and their organisational needs in design and development is also writ large in the AIA:

> "In eliminating or reducing risks related to the use of the high-risk AI system, due consideration shall be given to the technical knowledge, experience, education, training to be expected by the user and the environment in which the system is intended to be used." (AIA, Article 9)

The AIA also seeks to "encourage and facilitate" the uptake of voluntary codes of conduct to foster "stakeholders participation in the design and development of the AI systems" along with sustainability and accessibility requirements (AIA, Article 61). Whether mandated by name or not, the human oversight requirement reaches beyond a technical concern with interpretation and explanation to situate design and development in broader **human-centered processes and methods** that are the hallmark of HCI. It is not sufficient to design and develop interfaces that instruct users how to use AI systems, those interfaces must fit in with and support the organised 'doings' the user and the system are embedded in and enable users to take action to prevent or minimise harm to human health, safety, or fundamental rights in that context. The requirement to contextualise AI design and development – to give due consideration to the user and their environment – opens up AI to HCI and its wealth of experience to ensure AI systems are both practically and legally fit for purpose.

At the current moment in time, it appears as though the users of HRAIS are envisaged by the AIA to be persons acting in a professional capacity only.

> "For the purpose of this Regulation, the following definitions apply ... ... ... 'user' means any natural or legal person, public authority, agency or other body using an AI system under its authority, except where the AI system is used in the course of a personal non-professional activity." (AIA, Article 3)

While this is not an impediment to HCI, indeed it very much plays to it strengths in studying work, organisations and computer-supported cooperative work, it has led technology law scholars to suggest the term 'deployers' should be used instead of users [13]. More importantly, the lack of reference to non-professional users of HRAIS leaves a gap in the scope of the AIA with requiring oversight by consumers or citizens [4]. Indeed, it has been argued this group have no rights in the AIA and lack means to engage with approval processes or to influence whether or not products should be allowed onto the market in the first place [13]. It is worth noting that unlike GDPR, which confers fundamental rights on citizens, the AIA mainly focuses on product safety requirements across the supply chain and mediating market access. It is not a matter of argument that the AIA does not confer rights on citizens, it simply does not. Whilst this is a gap, many HRAIS systems will likely not only be subject to the AIA but also other regulatory regimes which do provide more direct citizen protections such as equality, human rights, data protection, and consumer protection laws. Furthermore, there is a strong and enduring argument to be made for including citizens in *responsible innovation* [60] processes to ensure the design and development of AI systems (high-risk or not) that are socially desirable as well as legally trustworthy, and this too plays to HCI expertise in co-design.

## 6 CONCLUSION

The law is often aligned with the encoding of rules in design [34], which has given rise to the idea that "code is law" or that it is possible to effect regulation through programming [39] and the "architecture of the internet" [48]. This is not the only possible relationship, however. The law is not simply a body of rules, but much more importantly a lively interpretive framework in which rules are invoked and their applicability contested. There is more than one possible relationship between law and design and we have sought to unpack here how proposed regulation of AI opens up possibilities for HCI that go beyond established concerns with the ethics of AI and explanation and what it means to design for the 'human in the loop' from a legal perspective. We have thus presented a number of legal provocations based on the introduction of common mandatory design and development requirements by the proposed European AI Act (AIA) for high-risk AI systems (HRAIS) that present a risk to human health, safety, and fundamental rights (such as privacy). These requirements include:

- Designing and developing data sets in compliance with **quality criteria and data governance** practices that regulate the training, validation and testing of data sets and commensurate development of models to minimise the risk of algorithmic discrimination. This requirement opens up possibilities for HCI in exploring the role of human-data interaction in federated machine learning systems.

- Testing AI systems for **accuracy**, consistent performance with respect to **intended use**, and the **identification of risks**, including risks occasioned by reasonably foreseeable misuse of the data, the model, or the system as a whole. This requirement opens up possibilities for HCI in developing scenarios for adversarial testing, the prospective identification and assessment of risk through usability testing, and retrospective monitoring of risk through post-deployment field studies.

- The design and development of novel forms of **technical documentation** supporting accountability and the demonstration of compliance. This opens up possibilities for HCI in co-developing human-centred architectures for AI systems in collaboration with AI developers and technology law scholars across different classes of autonomous system embedded within manifold application domains subject to manifold regulations.

- The design and development of **human-machine interface tools** that include but extend beyond methods of interpretation and explanation to the design of intelligent interfaces and collaborative decision-making solutions developed through contextually-sensitive, human-centred design methods and processes that enable users to oversee systems in use and **take action** to prevent or minimise risks to health, safety or fundamental rights.

We are aware of efforts to align HCI and AI [e.g., 71, 30, 56], that the relationship between the two fields is being explored through dedicated conferences including, for example, the International Conference on AI in HCI and the Annual Conference on Intelligent User Interfaces, and that HCI & AI courses proliferate. However, in articulating the various possibilities for HCI in the design of AI systems, this paper should not be read as sitting at the intersection between HCI and AI. We are not treading familiar ground even though we have hopefully touched upon it in places. Nonetheless, and distinctively, our interests lie in a different direction at the intersection between HCI and the **regulation** of AI. We have thus sought to consider how new rules put HCI at the front and centre of regulation, particularly with respect to human oversight and the need to situate the design and development of human-machine interface tools in context.

HCI has a lead role to play in the design and development of trustworthy AI systems that prevent or minimise risks to human health, safety, and fundamental rights and it is in this respect that two further provocations are occasioned by the legal imposition of design and development requirements. First, while it is clear that regulation anticipates a key role for human-centred design, AI systems development is currently dominated by ethical and technical agendas that shape the design process. This presents a fundamental challenge as to how legally mandated design and development requirements are to be *communicated within the design process* to shape trustworthy AI. This might be addressed through interdisciplinary collaboration between HCI researchers and legal scholars [e.g., 37]. Second, and perhaps more challenging, a further provocation lies in questioning the assumption that compliance with design and development requirements will actually produce trustworthy autonomous systems, as the functional steps specified in the AIA do not necessarily make significant contact with fundamental rights [23]. While health and safety requirements are covered by legislation relevant to a specific AI product, the possibility exists that a provider might put extensive resources into building a system that complies with design and development requirements but contravenes wider fundamental rights and is thus, by definition, untrustworthy. The design and development requirements of the AIA mandate what are basically good engineering practices around high-risk AI systems. They do not translate what designers need to do to comply with the largely non-functional requirements of fundamental rights and what steps developers need to take to protect them. This opens up the question of the kind of tools needed to ground the design and development of trustworthy autonomous systems in fundamental rights and the potential role of HCI in *assessing the impact of AI systems on fundamental rights*.

Seen from a regulatory perspective, and as elaborated in our consideration of legal provocations furnished by the AIA, HCI has a central role to play in ensuring that AI can comply with mandatory design and development requirements. These legal requirements make it clear that there is a lot more to the "human side" of AI than making methods of interpretation and explanation usable by non-developers. The legal provocations presented here not only project a potential role for HCI in the technical heartlands of AI and novel explorations of the accountability requirements mandated by technical documentation, but a critical role in developing human-machine interface tools to support human oversight. This is not only a matter of enabling users to interpret and understand how the machine works but also, and distinctively, of enabling users to take the situationally-relevant and contextually-appropriate actions needed to accomplish the organised 'doings' they are tasked with and to prevent or minimise risks to human health, safety, and fundamental rights in the process. The requirement that due consideration be given to the user and their environment creates new opportunities for HCI in AI systems development and mandates a fundamental turn to human-centred design methods and processes. The turn to HCI also projects further key roles for the community including developing means to communicate mandatory design and development requirements within the design process and assessing the impact of AI systems on fundamental human rights. Ultimately, the introduction of 'by design' regulation presents significant challenges to the providers of AI systems to ensure they are engineered around users and the environments in which they operate and that human-machine interface tools allow them to oversee the risks occasioned by the use of AI systems therein. In doing so, the regulation of AI opens up its design and development to interdisciplinary challenges that turn upon HCI's unique expertise. Thus, and to borrow once more from Harper [30], we might say *"the future is not AI; it can only be an AI enabled through HCI."*

## ACKNOWLEDGMENTS

This work was supported by the UK Engineering and Physical Sciences Research Council [grant numbers EP/V026607/1, EP/T022493/1] and UK Economic and Social Research Council [grant number ES/T00696X/1].

**Data Access Statement.** No research data is associated with this paper.